\documentclass[aps,twocolumn,showpacs,amsmath,amssymb,superscriptaddress,longbibliography]{revtex4-1} 
\usepackage{graphicx}
\usepackage{dcolumn}
\usepackage{color}
\usepackage{amsmath}
\usepackage[breaklinks,colorlinks,bookmarks=false,citecolor=green,linkcolor=blue,urlcolor=red]{hyperref}
\usepackage{braket}

\begin{document}
\title[J.\ Vahedi]{Edge magnetic properties of black phosphorene nanoribbons}

\author{Javad Vahedi}
   \email{j.vahediaghmashhadi@tu-braunschweig.de}
 \affiliation{Technische Universit\"{a}t Braunschweig, Institut f\"{u}r Mathematische Physik, Mendelssohnstra\ss e 3, 38106 Braunschweig, Germany}
\affiliation{Laboratoire de Physique Th\'eorique et Mod\'elisation, CNRS UMR 8089, CY Cergy Paris Universit\'e, 95302 Cergy-Pontoise Cedex, France.}

\affiliation{Department of Physics, Sari Branch, Islamic Azad University, Sari 48164-194, Iran}

\author{Robert Peters}
\email{peters@scphys.kyoto-u.ac.jp}
\affiliation{Department of Physics, Kyoto University, Kyoto 606-8502, Japan.}  

\date{\today}
\begin{abstract}
The magnetic properties of black phosphorene nanoribbons are investigated using static and dynamical mean-field theory.  Besides confirming the existence of ferromagnetic/antiferromagnetic edge magnetism, our detailed calculations using large unit-cells find a phase-transition at weak interaction strength to an incommensurate (IC) magnetic phase. A detailed Fourier analysis of the magnetization patterns in the  IC phase shows the existence of a second critical interaction strength, where the incommensurate phase changes to an antiferromagnetic (AFM) or ferromagnetic (FM) phase. We demonstrate that the difference of the ground state energies of the AFM and FM phase is exponentially small, making it possible to switch between both states by a small external field. Finally, we analyze the influence of strain and disorder on the magnetic properties and show that while the IC phase is robust to Anderson type disorder, it is fragile against strain.
\end{abstract}

\maketitle
\section{Introduction}
\label{sec1}
Phosphorene,  a novel promising 2D material, has recently attracted much attention owing to its anisotropic bandstructure~\cite{Liu14,Gomez16,Koenig14}. It is a bilayer puckered honeycomb lattice of black phosphorus with a peculiar bandstructure exhibiting Dirac cones in the bulk. Because of its bandstructure, phosphorene has been studied in many theoretical works, particularly in the context of transport studies~\cite{Ghosh17,Wang15,Linder17,Zare17}. Compared to the transition metal dichalcogenide materials, phosphorene has a high charge carrier mobility ($\sim 100~\text{cm}^2/\text{Vs}$) at room temperature~\cite{Liu14}, making it favorable for electronic applications. Moreover, zigzag phosphorene nanoribbons (ZPNR) exhibit two quasi-flat edge states, which are completely isolated from the bulk states~\cite{Ezawa14,Taghizadeh15,Ma16}, in contrast to the other 2D hexagonal lattice structures such as graphene~\cite{Castro09} and silicene~\cite{Shakouri15}. The nature of these isolated edge states originating from a large hopping parameter between two out-of-plane zigzag chains has been discussed in Ref.~\cite{Ezawa14}. Furthermore, a recent study addressed the Ruderman-Kittel-Kasuya-Yosida (RKKY) exchange interaction in ZPNRs. It found two different characteristic periods of the RKKY interaction mediating the magnetic interaction between impurities~\cite{Islam18}. 
\par
 Motivated by theoretical predictions~\cite{Yazyev10,Honecker10,Feldner11} and experimental
 confirmations~\cite{Magda14} of edge magnetism in zigzag graphene nanoribbons, edge magnetism has also been explored in 
 phosphorene in Ref.~\cite{Zhu14}. This first study found that ZPNRs display a magnetic state at the edge in the absence of a Peierls distortion. However, edge magnetism vanished in the fully relaxed structure. On the other hand, the authors of Ref.~\cite{Du15} have shown that edge magnetism of ZPNR, can survive even with structural relaxation.
 In another paper, the authors consider tilted black phosphorene nanoribbons (TPNRs) exposed to an external electric field. They found that the magnetic ground state can be switched by an electric field from antiferromagnetic (AFM) to ferromagnetic (FM)~\cite{Farooq16}. Furthermore, a quantum Monte Carlo calculation demonstrated a high Curie temperature for edge magnetism of ZPNR~\cite{Yang16}. 
However, previous studies have mainly considered small unit cells focusing on commensurate FM and AFM states. Magnetic states such as spiral phases or incommensurate phases have not been analyzed. Furthermore, the effect of strain and disorder on the magnetic states still remains unclear.
\par    
In this work, using a tight-binding (TB) Hubbard model, we numerically study the edge magnetism of ZPNRs using static mean-field theory (MFT) and dynamical mean-field theory (DMFT). Although QMC must be considered superior to our mean-field approaches, our (D)MFT is much faster and thus makes it possible to analyze large unit cells and incommensurate magnetic phases. 
Furthermore, mean-field theories have proven to at least qualitatively, sometimes even quantitatively, correctly describe magnetism in hexagonal 2D systems\cite{Marcin2020}, although being numerically less expensive. A similar combination of techniques has been used to analyze edge magnetism in zigzag graphene nanoribbons~\cite{Palacios07} and nanodots~\cite{Bhowmick08}. 
\par
In this paper, we demonstrate the existence of an incommensurate magnetic phase at the edge of ZPNR for weak interaction strengths $U_{c1}\lesssim U\lesssim U_{c2}$.
With increasing interaction strength, this incommensurate magnetic phase undergoes a phase transition into the ferromagnetic or antiferromagnetic phase at $U_{c2}$, which has been reported by previous studies. 
We show that the difference in the ground state energies of these two states is exponentially small, making it easy to switch between both states. Besides, to gain more insight into the realization of magnetism in ZPNRs at weak interaction strengths, the purpose of this paper is to analyze the effects of strain and defects on the magnetic state. 
Such perturbations of the material are ubiquitous in  2D materials~\cite{Gui08,Wang14,Yue12,Tabatabaei13,Rodin14,Elahi15,Fazileh16}. Moreover, studies on strain in nonmagnetic phosphorene show some intriguing features: A first-principle study predicted a semiconductor-semimetal-metal transition under perpendicular compression~\cite{Rodin14}. In Ref.~\cite{Elahi15}, an emergence of a peculiar Dirac-shaped dispersion for tensile strain in the zigzag edge is proposed.  In another work, it was shown that tensile or in-plane strain, together with spin-orbit interaction, gives rise to a topological phase transition~ \cite{Fazileh16}. However, the only study which analyzes the effect of strain on ZPNRs magnetism is found in Ref.~\cite{Du15}. It predicts that at a critical compressive strain along the zigzag edge (about 5$\%$), the ground state changes from an AFM semiconductor to a nonmagnetic metal. 
Thus, we here address the effect of strain and Anderson type disorder on the magnetic properties of ZPNRs  and find that while the IC phase is very sensitive to strain and disappears fast,  it is robust against Anderson type disorder. We also notice that the second critical point $U_{c2}$ shifts to larger values. Thus,  one can predict that the AFM/FM magnetic phase disappears under large strain.
\par
The paper is organized as follows: In Sec.\ \ref{sec2},  we introduce the theoretical model and formalism used in the numerical calculations. In Sec.\ \ref{sec3}, we discuss results obtained. Finally, we summarize and conclude our results in Sec.\
  \ref{sec4}.
\begin{figure}[b]
\includegraphics[width=1\columnwidth]{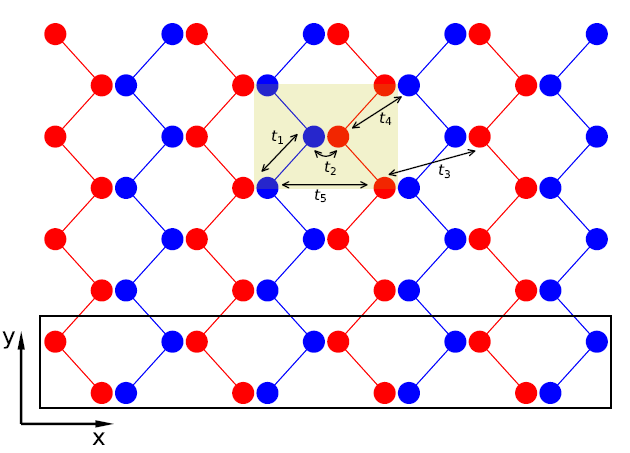}
\caption{
A schematic view of the ZPNR: The yellow area corresponds to the unit cell, which consists of four atoms. The black arrows show the hoppings up to the fifth nearest-neighbor hopping included in our TB model. The red (blue) circles indicate the upper (lower) layers. The black box is the ribbon unit cell in the y-direction. In the text, $N_y$ refers to the number of ribbon unit cells in the y-direction. The width of the unit cell is specified by $N_x$, which includes $N=4\times N_x$ phosphorus atoms.}
 \label{fig1}
\end{figure}
\section{MODEL AND FORMALISM}
\label{sec2}
In order to study  the magnetic properties of ZPNRs, we use the following tight-binding model
\begin{equation}
 H=\sum_{ij,\sigma}t_{ij}c_{i\sigma}^{\dagger}c_{j\sigma}+U\, \sum_i \left(n_{i,\uparrow}-\frac{1}{2}\right)\,\left(n_{i,\downarrow} -\frac{1}{2}\right),
 \label{eq1}
\end{equation}
where the first summation runs up to the fifth nearest neighbor, and $t_{ij}$ is the hopping integral proposed in Ref.~\cite{Rudenko14}.  These hopping parameters are $t_1 = -1.220~eV$, $t_2 = 3.665~eV$, $t_3 =-0.205~eV$, $t_4 = -0.105~eV$, and $t_5 = -0.055~eV$, which are shown by arrows in Fig.~\ref{fig1}. Furthermore, we include a local density-density interaction. Thus, this model corresponds to  a one-band Hubbard model. To tackle the interaction, we use static and dynamical mean-field theory.  At the static mean-field level (MFT) level, all quantum fluctuations are neglected, and the SU(2) spin symmetry must be broken artificially in order to capture the formation of local moments.  An extension of the MFT to account for the local moment formation is the dynamical mean-field theory (DMFT). The DMFT approximation accounts for temporal fluctuations and thus includes local charge fluctuations beyond static MFT. Indeed, the accuracy of DMFT to predict the critical point, $U_c$, in the honeycomb lattice  has been reported recently in Ref.\cite{Marcin2020,Thu2020}.

\subsection{Static mean-field theory (MFT)}
Evaluating the Coulomb interaction term  in the mean-field approximation leads to two potentials terms, direct and exchange term, which must be solved self-consistently. In the case of the Hubbard model in the collinear approximation, only the
direct potential term is nonzero, and one obtains
$$
 U\,\sum_i \Bigl( \langle n_{i,\uparrow} \rangle n_{i,\downarrow} + n_{i,\uparrow}\langle n_{i,\downarrow}\rangle
- \langle n_{i,\uparrow}\rangle \langle n_{i,\downarrow}\rangle-\frac{n_{i,\uparrow}+ n_{i,\downarrow}}{2}
+ \frac{1}{4} \Bigr)
$$
where $n_{i\sigma}=c_{i\sigma}^{\dagger}c_{i\sigma}$ is the number operator and $\langle  n_{i\sigma}\rangle$ is the average electron occupation number for spin-down $(\downarrow)$ and spin-up $(\uparrow)$ electrons on lattice site $i$. We focus on the undoped ZPNR with exactly one electron per lattice site, i.e., we work with the half-filled Hubbard model. 
\par
To calculate the magnetic ground state of Hamiltonian Eq.~(\ref{eq1}), we start with a few initial, specific or random, configurations for the average electron occupation number $\langle  n_{i\sigma}\rangle$. Then, by diagonalizing the Hamiltonian, we calculate updated  electron occupation numbers. This procedure is repeated until the convergence criteria, chosen as $\eta=10^{-8}$, is achieved on the average electron occupation number. This self-consistent solution provides the local magnetization $m_i^z=(n_{i\uparrow}-n_{i\downarrow})/2$ on each site. Finally, the energies of different states are compared to find the ground state.

\subsection{Dynamical mean-field theory (DMFT)}
A recent study has shown that the transition to the magnetic state in Graphene is captured remarkably well by the  inclusion of local charge fluctuations ~\cite{Marcin2020} in the framework of  a single-site dynamical mean-field theory~\cite{Georges1996}.
Thus, to go beyond the static MFT and to include local fluctuations, we also use the real-space dynamical mean-field theory (DMFT) to obtain a magnetic solution of the ZPNR.
As in Ref. \cite{Thu2020}, each atom of a $8\times 48$ large cluster is mapped onto its own quantum impurity model by calculating the Green's function and the local hybridization function,
\begin{eqnarray}
\bf{G}_{ij}(\omega)&=&(\omega-\bf{H}-\bf{\Sigma}(\omega))_{ij}^{-1}\\
\bf{\Delta}_{i}(\omega)&=&\bf{G}_{ii}^{-1}(\omega)+\bf{\Sigma}_{ii}(\omega),
\end{eqnarray}
where $i$ and $j$ are indices for the positions of the atoms, $\bf{H}$ is the matrix of the tight-binding Hamiltonian on the finite lattice, $\bf{\Sigma}(\omega)$ is the diagonal matrix including the self-energies of all atoms $\bf{\Sigma}_{ij}(\omega)=\bf{\Sigma}_{ij}(\omega)\delta_{ij}$, $\bf{\Sigma}_{ii}$ is the self-energy of the atom $i$, $\bf{G}_{ij}$ is the Green's function matrix, and $\bf{\Delta}_i$ is the hybridization function of atom $i$. The hybridization function together with the local interaction strength completely defines a quantum impurity model necessary in DMFT, which makes it possible to calculate magnetic states in large clusters\cite{PhysRevB.89.155134,PhysRevB.92.075103,Marcin2020,Thu2020}.
The quantum impurity model is then solved using the numerical renormalization group (NRG)\cite{RevModPhys.47.773,RevModPhys.80.395}, which can calculate dynamical correlation functions and self-energies with high accuracy\cite{PhysRevB.74.245114}.
\begin{figure}
\includegraphics[width=1\columnwidth]{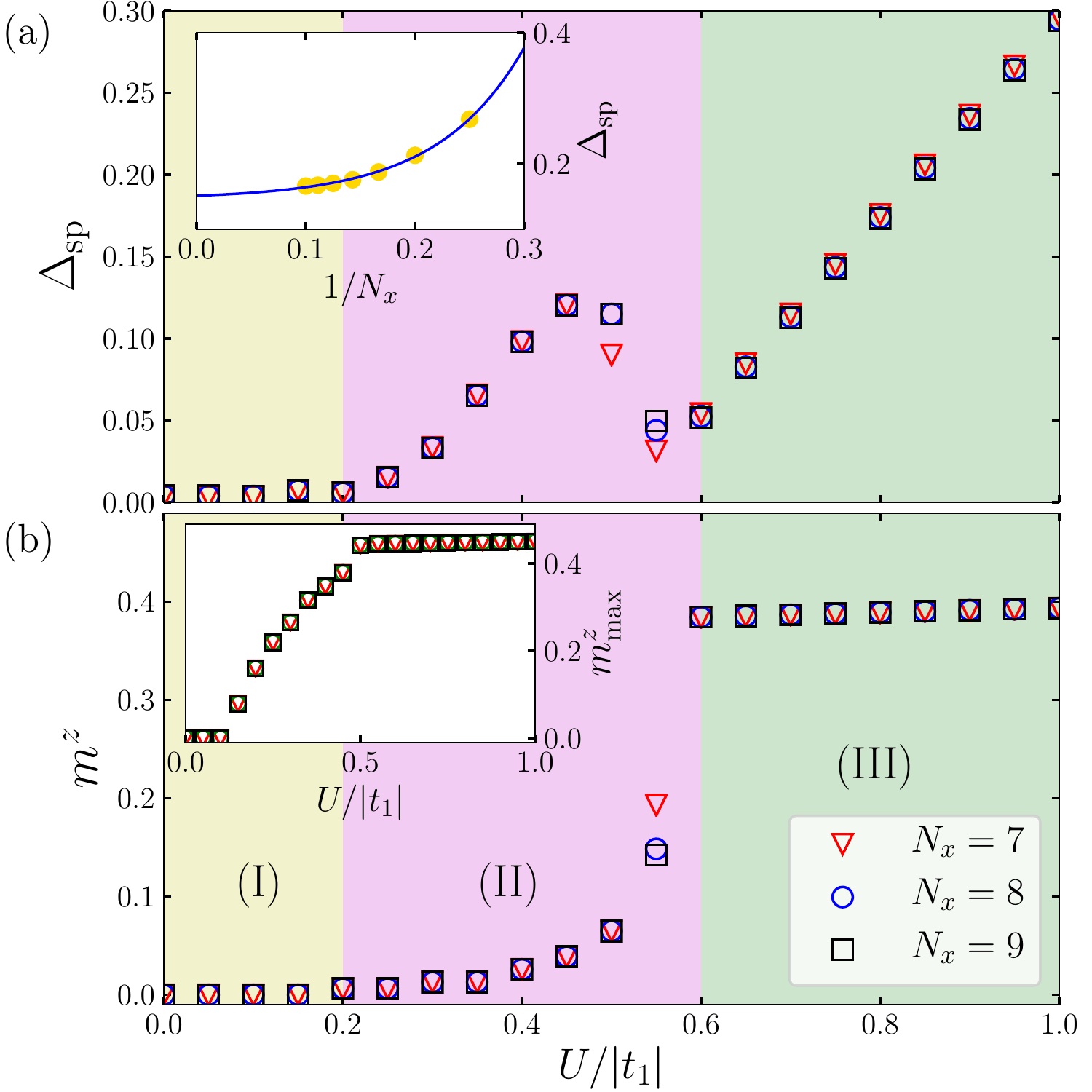}
\caption{
Evolution of the single-particle gap (a) and the edge  magnetization (b) as a function of the Hubbard interaction $U/|t_1|$. Results for different ZPNR widths $N_x=7,8,9$ are plotted with different symbols.  In panel (a),  the inset shows an exponential curve-fitting of the gap evolution as a function  of $1/N_x$ at $U/|t_1|=0.8$. In panel (b),  the inset shows the evolution of the maximum magnetization $m^z_{\rm max}$ versus the Hubbard interaction.  Three different regimes are labeled and highlighted as: (I)  nonmagnetic, (II) gapped-IC, and (III) gapped-AFM (or gapped-FM) regions. The ribbon length is fixed at $N_y=120$, and the periodic boundary condition is implemented in the $y$-direction.}
\label{fig2}
\end{figure}
\begin{figure}[t]
\includegraphics[width=1\columnwidth]{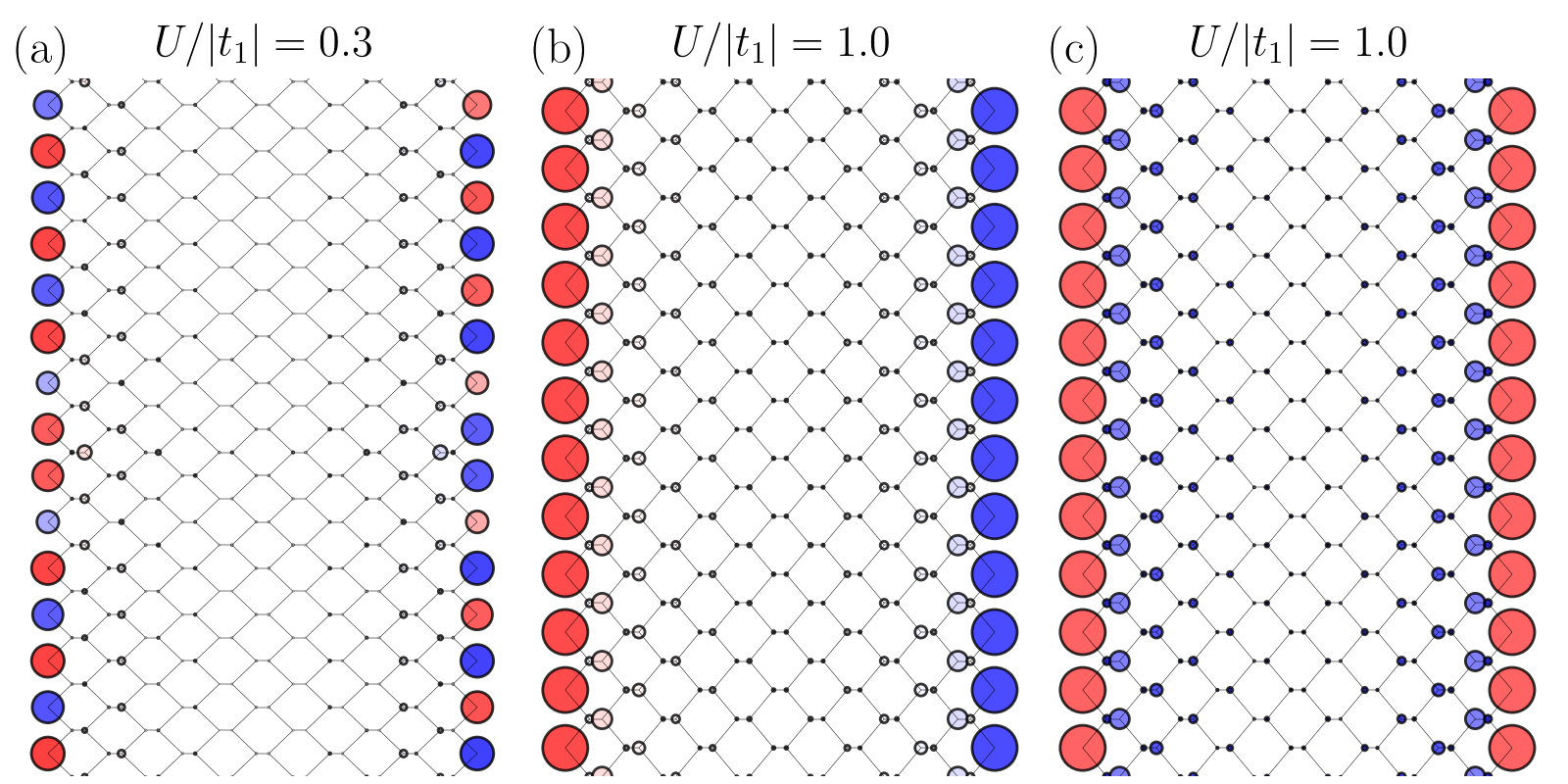}
\caption{
Three different magnetic spatial configurations for $U/|t_1|=0.3$ (left panel) and $1.0$ (middle and right panels) of the Hubbard interaction. The width and length of the ribbons are $N_x=7$ and $N_y=120$. Since we here use a very long ribbon, we only show a portion of the ribbon in the y-direction. The blue and red circles display the two different local spin directions. We call the magnetic configuration in the left, middle, and right panel as IC, AFM, and FM phases.}
\label{fig3}
\end{figure}
\begin{figure*}
\includegraphics[width=2\columnwidth]{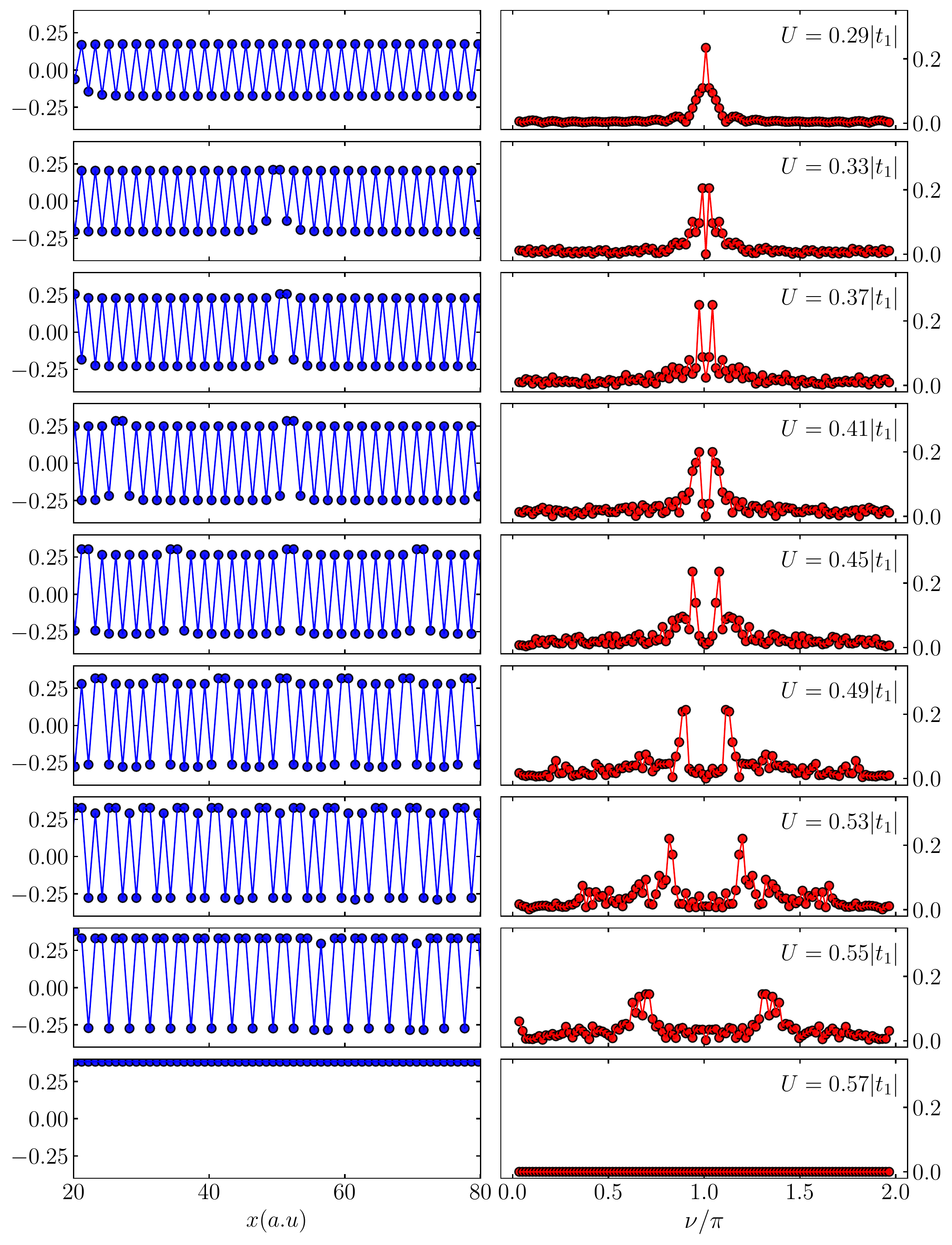} 
\caption{
The left panels present the local magnetic modulation $m_i^z$ at one edge of the ZPNRs, and the right panels give their corresponding Fourier transformation for different Hubbard interactions $U/|t_1|$. The lattice parameters are the same as in Fig.~\ref{fig2}.}
\label{fig4}
\end{figure*}
\section{RESULTS AND DISCUSSION}
\label{sec3}
In this section, we present the numerical results obtained by static and dynamical MFT.  The  ZPNR geometry is shown in Fig.~\ref{fig1}. The geometry is specified by two parameters, $N_x$ and $N_y$, which are the width and the length of the cluster. We use open boundaries in the x-direction. A single ZPNR unit-cell is shown as a black box in  Fig.~\ref{fig1}. For  our static MFT calculations, we use $N_y=120$ unit-cells in y-direction and apply periodic boundary conditions. Thus, our calculation includes  $N=4\times N_x\times N_y$ phosphorus atoms.  The ribbon width plays an essential role in the creation of the edge states~\cite{Taghizadeh15}, and it has been shown that the ribbon width must be larger than about $3~nm$  for stable edge magnetism, which corresponds to $N_x=7$ in this work.
\subsection{The pristine ZPNRs}
We first consider a finite ribbon cluster. Later, we exploit translation symmetry and extend our study to an infinite ribbon. We focus here on the single-particle gap and the edge magnetization in our analysis, which are two practical observables to understand the magnetic features of ZPNRs. The single-particle gap is here defined as one half of the charge
gap, $\Delta_{\rm sp} = (E_{n-1} - 2\, E_{n} + E_{n+1})/2$, where $E_n$ is the ground-state energy in the sector with $n$ electrons. The  edge magnetization is defined  as $m^z=\frac{1}{N_{\rm edge}}\sum_{i\in {\rm edge}}^{N_{\rm edge}}\, |\langle m_i^z \rangle|$. The temperature is set to zero. 
\par
Figure~\ref{fig2} shows the evolution of the single-particle gap (a) and the edge magnetizations (b) as a function of the Hubbard interaction $U/|t_1|$, calculated by static MFT.  The gap is zero for interaction strengths $U<U_{c1}\simeq0.2|t_1|$. At this point, the edge magnetism starts to appear. For $0.2\lesssim U/|t_1|\lesssim0.6$, the magnetization at the edge is not homogeneous. Precisely at the critical point, $U_{c1}$, the magnetization pattern of one edge is an antiferromagnetic state, whose existence has been reported in Ref.~\cite{Du15}. Further increasing the interaction strength, the magnetic state becomes an incommensurate (IC) antiferromagnetic state (see the left panel in Fig.~\ref{fig3} and Fig.~\ref{fig4}). A more detailed analysis is given in the next section. 
 As can be seen from Fig.~\ref{fig2}(a), the gap starts to increase from the first critical point $U_{c1}\simeq0.2|t_1|$. Surprisingly, the band gap forms a cusp, and decreases for stronger interaction strengths until it reaches the second critical point $U_{c2}\simeq0.6|t_1|$. Beyond $U_{c2}\simeq0.6|t_1|$, the gap shows a linear growth with the Hubbard interaction. The magnetic configuration in this region is illustrated in the middle panel of Fig.~\ref{fig3}, and we refer to it as the AFM phase. 
 While the magnetization along the edges is homogeneous, the magnetization is exactly opposite at both edges.
  This configuration has also been predicted by DFT~\cite{Zhu14} and QMC~\cite{Yang16}. However, besides this AFM phase, we here find another magnetic solution illustrated in the right panel of Fig.~\ref{fig3}. In this magnetic configuration, both edges are ferromagnetically aligned, and interestingly its gap and magnetization behavior are almost the same as in the AFM case.
    Comparing the ground state energies of the AFM and the FM states, we find an exponentially small energy difference of $\mathcal{O}(10^{-6})$ (see Table.\ref{tab_EG}). It is worth mentioning that to find the AFM or the FM  states, a proper  initial guess is necessary while finding the IC phase does not require such an initial guess.  We highlight and label the ZPNRs phases in Fig.~\ref{fig2} as follows: (I) nonmagnetic, (II) gapped-IC, and (III) gapped-AFM  (or gapped-FM ) regions. 
  \par
  It is intriguing to see that an FM state in a half-filled Hubbard model has a slightly lower ground state energy than the AFM state. However, we might explain this ferromagnetic state by using the argumentation of Stoner ferromagnetism: The wavefunctions of the edge states of the left and the right edges have some overlap with each other if the width of the ZPNR is finite. This overlap will lead to an additional positive energy contribution in the case of an AFM state, which can be prevented by ferromagnetically aligning both edges. Thus, the ferromagnetic state has slightly lower energy. The situation could be  similar to graphene zigzag ribbons, where a sharp semiconductor (AF) to metallic (FM) transition occurs by varying the ribbon width, which is seen experimentally~\cite{Magda14} and theoretically~\cite{Chen17}.
\begin{table}[b]
\centering
\begin{tabular}{ |c|c|c|c|c| } 
\hline
$U/t_1$ & 0.8 & 0.9 & 1.0  \\
\hline
E[eV]/N& -3.1342389394 & -3.0761689920 & -3.0181486070\\ 
FM&  &  & \\
\hline
E[eV]/N& -3.1342358713 & -3.0761659724 & -3.0181456815\\
AF&  &  & \\
\hline
\end{tabular}
\caption{Ground-state energy per atom for the gapped-AFM  and the gapped-FM phase for three interaction strengths.}
\label{tab_EG}
\end{table}
\begin{figure}[t]
\includegraphics[width=1\columnwidth]{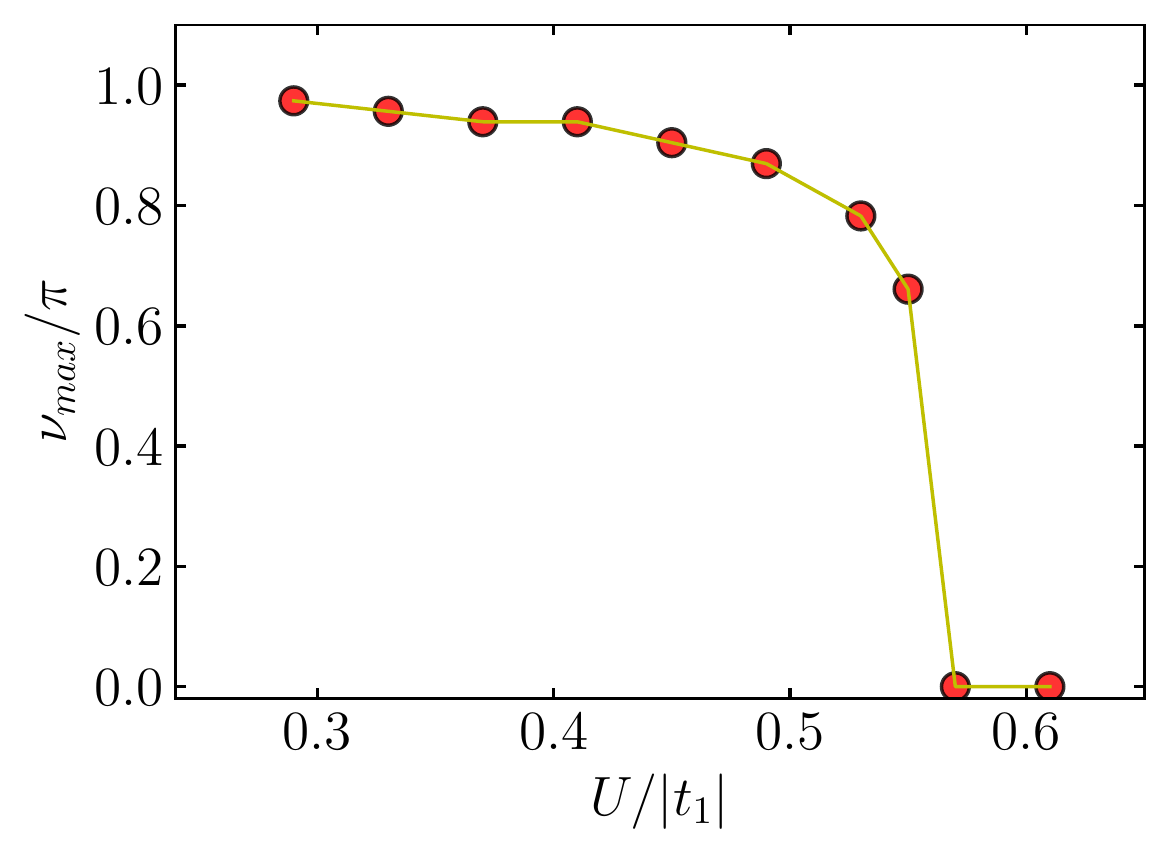} 
\caption{
The position of the maximum amplitude in the frequency domain as a function of the Hubbard interaction $U/|t_1|$ in Fig.~\ref{fig4}.}
\label{fig5}
\end{figure}
\begin{figure}[b]
\includegraphics[width=1\columnwidth]{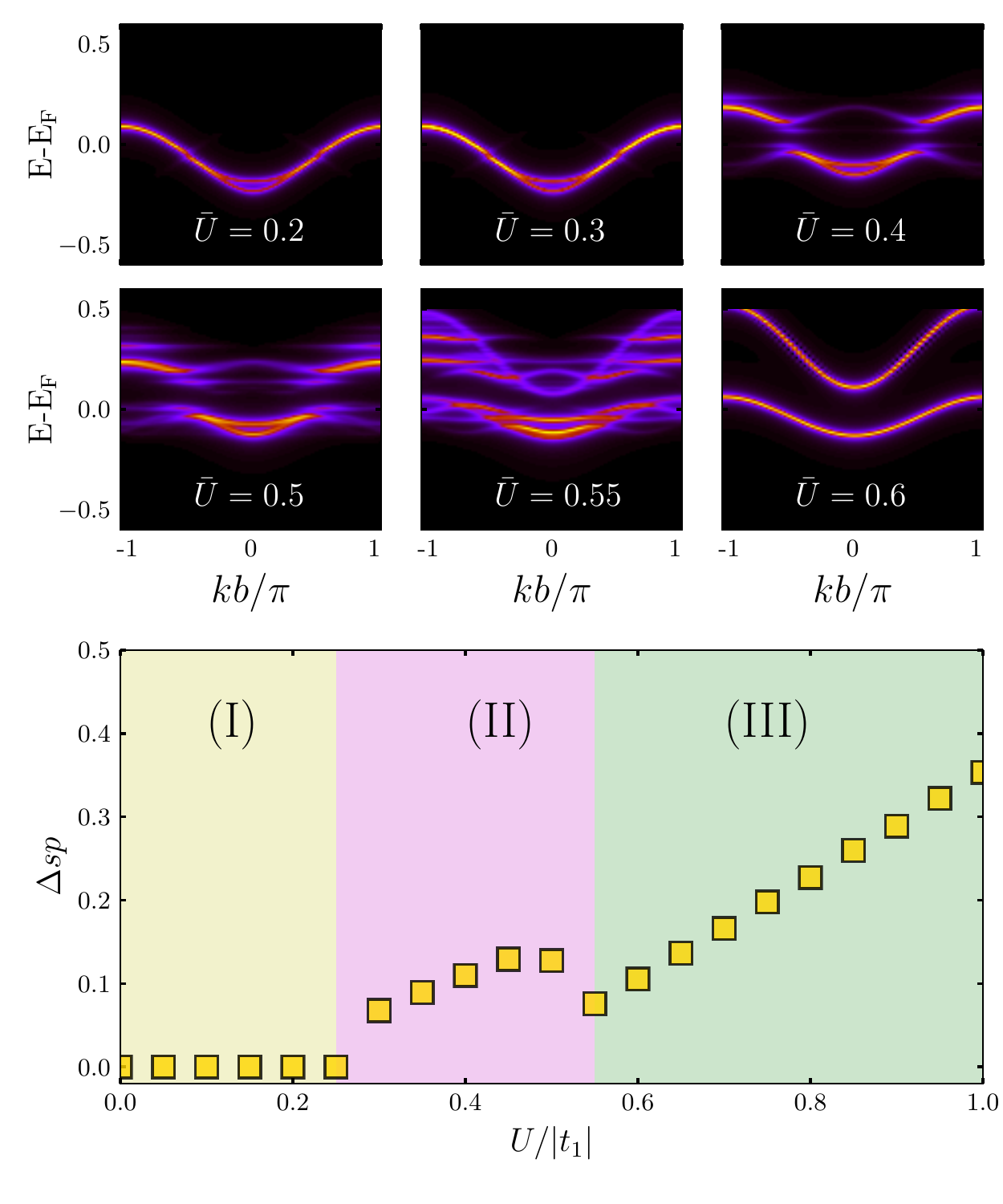} \\[0mm]
\caption{
Top and middle panels show spectral spectrum  which extracted by unfolding energy spectrum of extended unit-cell with $Ny=120$. The bottom panel shows the evolution of single-particle gaps. Regions are as follows: (I) nonmagnetic, (II) gapped-IC, and (III) gapped-AFM. The width of the ribbon is fixed at $N_x=7$. }
\label{fig6}
\end{figure}

\par
Figure~\ref{fig2}(b) shows the edge magnetization $m^z$. To calculate the edge magnetization, we only consider lattice sites along the border of the zigzag edge. The general behavior is consistent with the gap evolution. A  small magnetization  appears at the first critical point $U_{c1}\simeq0.2|t_1|$ and increases very slowly with $U$ until the second critical point $U_{c2}\simeq0.6|t_1|$. Beyond $U_{c2}$, the edge magnetization $m^z$ saturates. We note that our results of phase (III) are in agreement with the quantum Monte Carlo result reported in Ref.~\cite{Yang16}, which predicted long-range order for $U>0.5~eV$ at zero temperature. The inset of  Figs.~\ref{fig2}(b) shows the maximum of the edge magnetization $m^z_{\rm max}$. Interestingly,  it captures both the first and the second critical point consistent with the gap evolution. 
\par
To obtain more insight into the effect of the ribbon's width on the gap and the magnetization, we present in Fig.~\ref{fig2} data of different widths as a comparison. One can see  that the gap does not depend on the ribbon width for $N_x=7,8,9$, for which the data collapse on top of each other. However, for ribbon widths smaller than $N_x<7$, we find that the gap depends on the width. The inset in Fig.~\ref{fig2}(a) displays the gap evolution with the inverse ribbon width,  fitted by an exponential curve. We find that the edge magnetization, $m^z$, also collapses on a single curve for $N_x>6$. In particular, all data show the same saturation value in the region-(III).  However, we note that the QMC calculation~\cite{Yang16} for room temperature has shown that the magnetization decreases with the ribbon width.
\par
Let us now analyze the IC phase in more detail by accessing the real space data of the local magnetization, $m_i^z$.
The real-space data of the local magnetization, $m_i^z$, reveals how an antiferromagnetic state at one edge changes into the ferromagnetic state when increasing the interaction strength. The local magnetization and its Fourier transformation (FT) are shown in Figs.~\ref{fig4} for different interaction strengths. For small $U/|t_1|=0.25$, the local magnetization pattern is an antiferromagnetic state along one edge, as also demonstrated by the FT with a single peak at $\nu_{\rm max}=\pi$. By increasing $U$, one can see how the local magnetization starts to change.  The single peak in the FT splits into two, which move away from $\pi$. Finally, for a Hubbard interactions larger than $U>U_{c2}$, the maximum in the FT occurs at $\nu_{\rm max}=0$, which signals a fully aligned ferromagnetic state along one edge.
 In Fig~\ref{fig5}, we show the position of the maximum in the FT $\nu_{\rm max}/\pi$ plotted as function of $U/|t_1|$. It can be seen how the maximum decreases from $1$ to $0$ within the IC phase.
\begin{figure}[t]
\includegraphics[width=1\columnwidth]{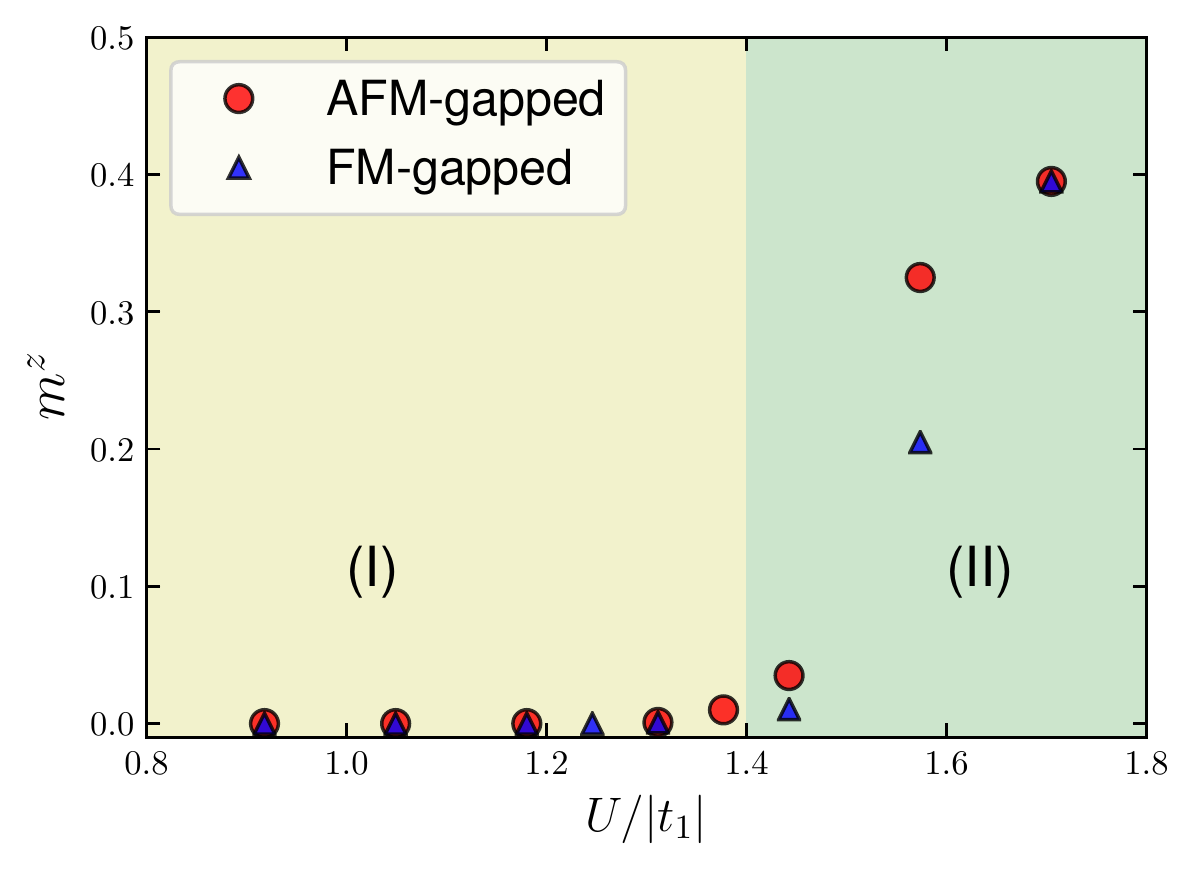} \\[0mm]
\includegraphics[width=1\columnwidth]{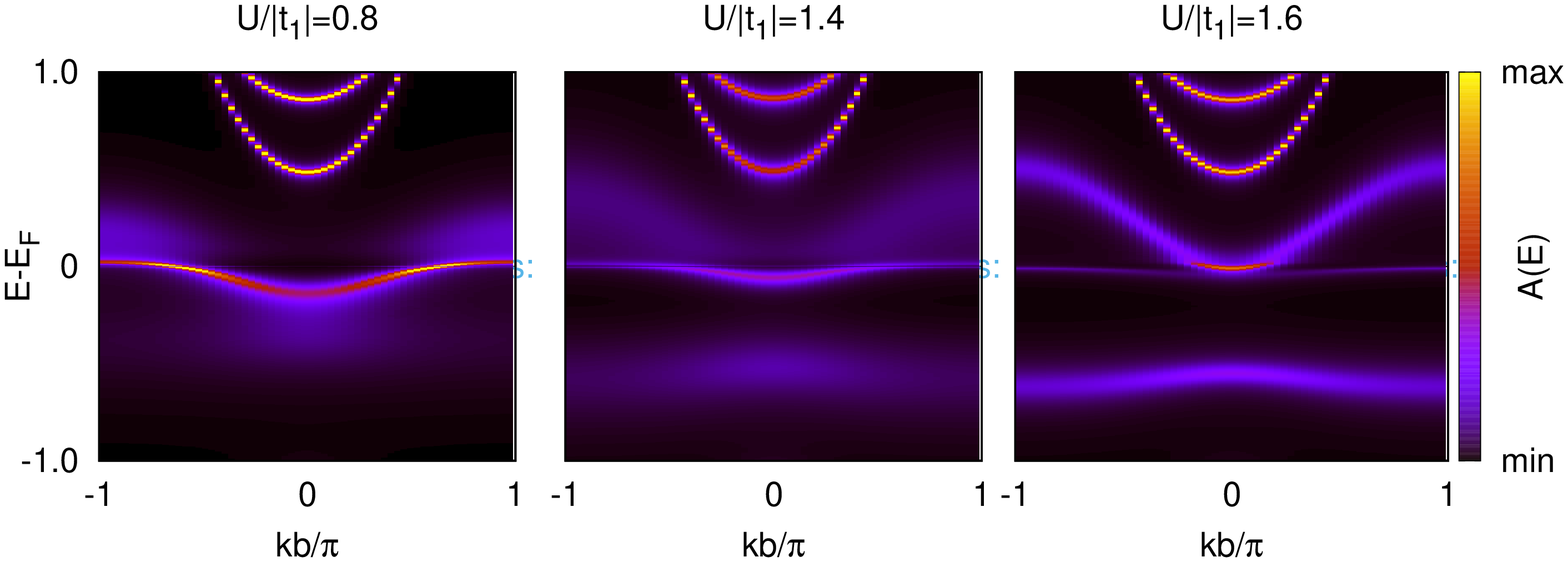} 
\caption{
DMFT calculations of edge  magnetization (top panel) as a function of Hubbard interaction $U/|t_1|$ for  a ZPNR width $N_x=8$.  Two different regimes are labeled and highlighted as: (I)  nonmagnetic,  and (II) AFM gapped (or FM gapped ) regions. Bottom panels show the spectral functions $A(E)$ for three different Hubbard interactions. The ribbon length is fixed at $N_y=48$ and the periodic boundary condition is implemented in the $y$-direction.}
\label{fig6b}
\end{figure}
\begin{figure}[b]
\includegraphics[width=1\columnwidth]{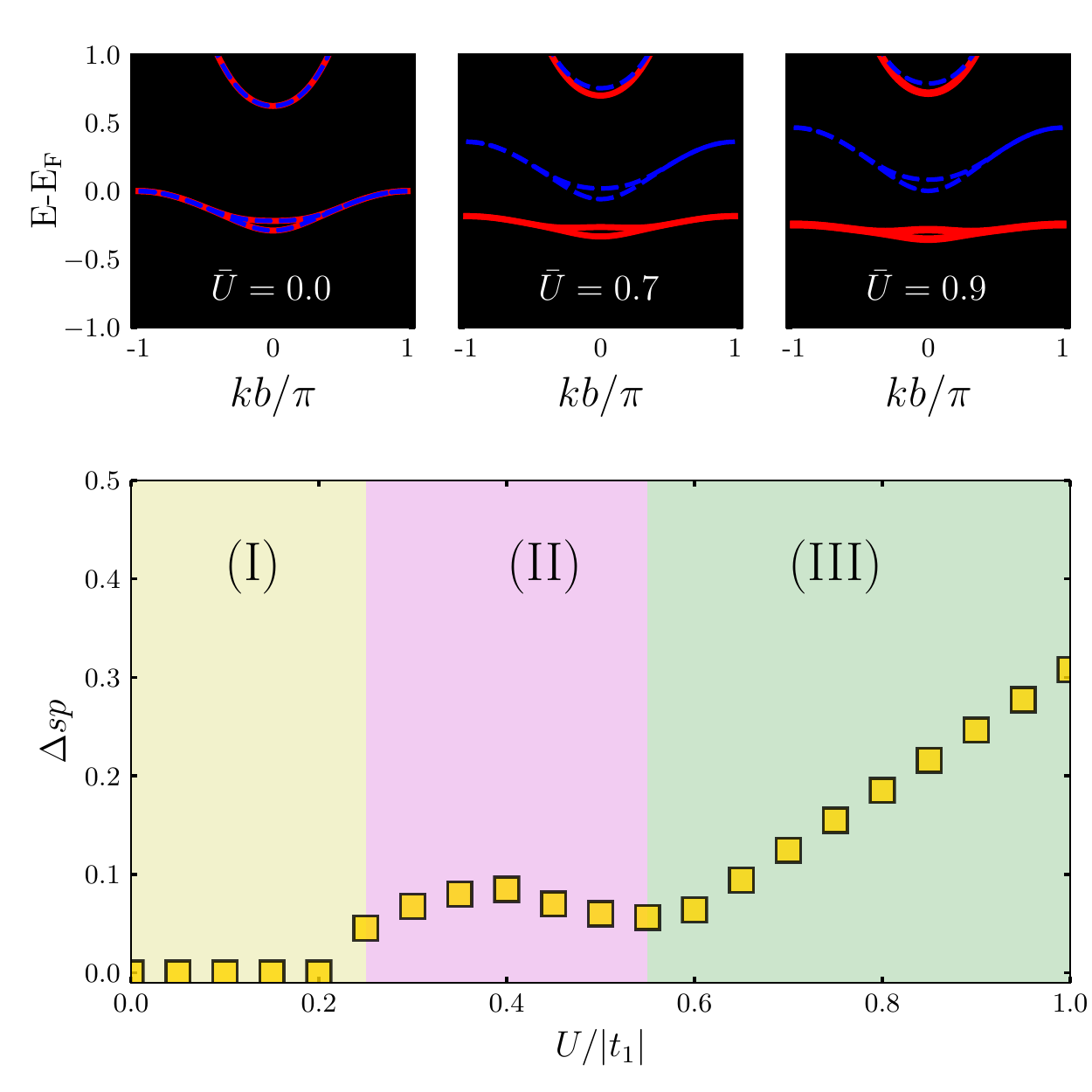} 
\caption{
Same as Fig.\ref{fig6}, but for the FM  case. Regions are as follows: (I) nonmagnetic, (II) gapped-IC, and (III) gapped-FM. Red and blue lines in the spectral functions correspond to different spin-directions.}
\label{fig7}
\end{figure}
\par
Now, we use the translational symmetry of the lattice and probe the magnetic features of an infinite ribbon. To this end, we focus on the energy dispersion of an infinite ribbon. For a given wavenumber $k$ and spin $\sigma$ the mean-field Hamiltonian has $N$ states $\Psi_{k\sigma}(x)$ with energy $\epsilon_{k\sigma}(x)$. By implementing the iterative self-consistent technique~\cite{Palacios07} in the Brillouin zone (BZ), we can recover the magnetization configuration in  ZPNRs. It worth to mention that the magnetic unit-cell in the IC phase is much larger than the lattice unit-cell. Thus, we use an unfolding technique to extract the spectral function. Detailed information about the unfolding is given in the appendix \ref{Unfold}.
\par
In Fig.~\ref{fig6}, the spectral function of the edge states and the corresponding gap evolution are illustrated. We show that the quasi-flat bands of the edges are isolated from the bulk bands, which is the most prominent feature of ZPNRs. It is already known that the hopping $t_4$ term is responsible for the dispersion of these flat bands~\cite{Ezawa14}. We also note that these quasi-flat modes are almost doubly degenerate.  As shown in Fig.~\ref{fig6} in the non-magnetic region  ($\bar{U}\equiv U/|t_1|=0.2$), these bands are degenerate at the BZ boundaries and split toward the BZ center. Indeed, this splitting becomes smaller for wider ribbons. The following reasoning may explain this: increasing the width of the ZPNRs will reduce the interaction between both edges, leading to the decrease of the edge splitting at the BZ center. When entering the magnetic phase ($\bar{U}\gtrsim0.2$), the bands split at the $kb=\pm\pi/2$ points at the Fermi energy. With increasing interaction strength, the gap size increases, and spectral weight is shifted at  $k=0$ above the Fermi energy and at $kb=\pm\pi$ below the Fermi energy. This shift of spectral weight causes the single-particle gap to decrease before entering the ferromagnetic phase (region-(III)). Finally, the spectral weight above and below the Fermi energy form two quasiparticle bands for $U>0.6$. In region-(III) with commensurate (here, gapped-AFM) edge magnetism, one band is shifted toward higher energies, and both bands are separated. The gap is clearly visible in this phase, and we can read off the gap in the energy dispersion. The gap evolution is thereby similar as in Fig.~\ref{fig2} directly calculated by the energy. We find a gap opening when entering the IC phase. The gap width forms a maximum in the IC phase, decreases towards the AFM phase, and finally increases linearly in the AFM phase.
\par
To further validate our MFT results, we will now show DMFT results. As mentioned above, DMFT has proven to predict the critical point in Graphene adequately when compared to lattice-QMC. We here use DMFT for a cluster with parameters $N_x=8$ and $N_y=48$. Our results are summarized in Fig. \ref{fig6b}. For weak interaction strengths, $0.7<U/\vert t_1\vert$, we find a nonmagnetic solution. We note that any (even a magnetic) initial guess for these interaction strengths converges to the same nonmagnetic state. Furthermore, we find a stable magnetic state at the edges of ZPNR for $U/\vert t_1\vert>1.4$, which corresponds to phase (III) in MFT. As with MFT, we can find a stable AFM and a stable FM state. As in Graphene, local fluctuations included by DMFT shift the critical point to stronger interaction strengths compared to static MFT.
More interestingly is the question about the existence of the IC phase. For interaction strengths $U/\vert t_1\vert<1.4$, we do not find a converged magnetic solution. However, for $0.7<U/\vert t_1\vert<1.4$, we find a small magnetization at the edges of ZPNR, which does not vanish when iterating the DMFT calculation. If we start the DMFT in this regime with an inhomogeneous magnetic state, the magnetization configuration changes in each iteration without completely vanishing, but we cannot find a converged solution. We note here that to find an incommensurate state with DMFT, a large cluster and an appropriate initial guess are necessary. Thus, we interpret these DMFT calculations as an attempt to stabilize an incommensurate state. However, as we cannot find a converged solution, we cannot calculate further properties of this phase.
\par
An advantage of DMFT over static MFT is that spectral functions can readily be calculated and include lifetime effects due to correlations. The spectral functions calculated by DMFT are shown in Fig. \ref{fig6b}. In contrast to static MFT, DMFT already includes modifications of the spectral function in the nonmagnetic phase (I). For $U/\vert t_1\vert=0.8$ (converged nonmagnetic solution), we find some (blurred) spectral weight below the quasiparticle band at $k=0$. This spectral weight should correspond to the splitting of the quasiparticle band at the center of the BZ, which is smeared out because of correlations. Furthermore, we find some spectral weight above the quasiparticle band at $kb=\pm\pi$. With increasing interaction strength, spectral weight is particularly transferred from the quasiparticle band lying at the Fermi energy to energies below the Fermi energy, slowly forming a second band. At the same time, the spectral weight moves to higher energies at $kb=\pm \pi$. These two processes finally form two bands, which are clearly visible for $U/\vert t_t\vert=1.6$ with a gap between them. While in static MFT, the formation of a gap takes place when entering the AFM phase (III), in DMFT this separation already starts in the nonmagnetic phase due to local fluctuations.
\par
Finally, we want to examine the above mentioned gapped-FM phase in more detail, which coexists with the gapped-AFM phase. We repeat the previous calculations, using a proper initial guess to obtain the FM phase.  The results are depicted in Fig.~\ref{fig7}. We note that when both edges are ferromagnetically aligned, we can also find the IC phase.
The gap evolution reveals a small {\em dome} in the region-(II).  The gap opens at the first critical point, $U_{c1}$, forms a maximum and then decreases when approaching the second critical point $U_{c2}$. In the region-(III), the gap increases linearly with the Hubbard strength interaction. It is interesting to note that the energy dispersion of the AFM state (Fig.~\ref{fig6}) and the FM state (Fig.~\ref{fig7}) are almost identical. However, while in the AFM state, all edge modes are spin-degenerate, in the FM state, the edge modes above the Fermi energy have a definite spin-directions and the edge modes below the Fermi energy exhibit an opposite spin direction. Furthermore, there is a slight additional splitting of the edge modes around $k=0$ in the FM state, which is absent in the AFM state. This small splitting is responsible for the energy difference between the AFM and the FM state.

\subsection{Strain effects}
Next, we want to study the impact of strain and disorder on the magnetic state.
To study strain effects, we follow the approach developed in Ref.~\cite{Jiang15, Mohammadi16}.  We
will focus here on the tensile strain in the normal direction to the phosphorene plane~\cite{Huang14}. By applying an
axial strain,  following the Harrison relation~\cite{Harrison04}, the strain-induced modified hopping parameter in the linear regime can be written as $t_i\approx(1-2\alpha_x^i\epsilon_x-2\alpha_y^i\epsilon_y-2\alpha_z^i\epsilon_z)$, where $\alpha_i^j$ are coefficients related to the structure of phosphorene and $\epsilon_j$  is the strain in the $j$-direction. 
\par
\begin{figure}
\includegraphics[width=1\columnwidth]{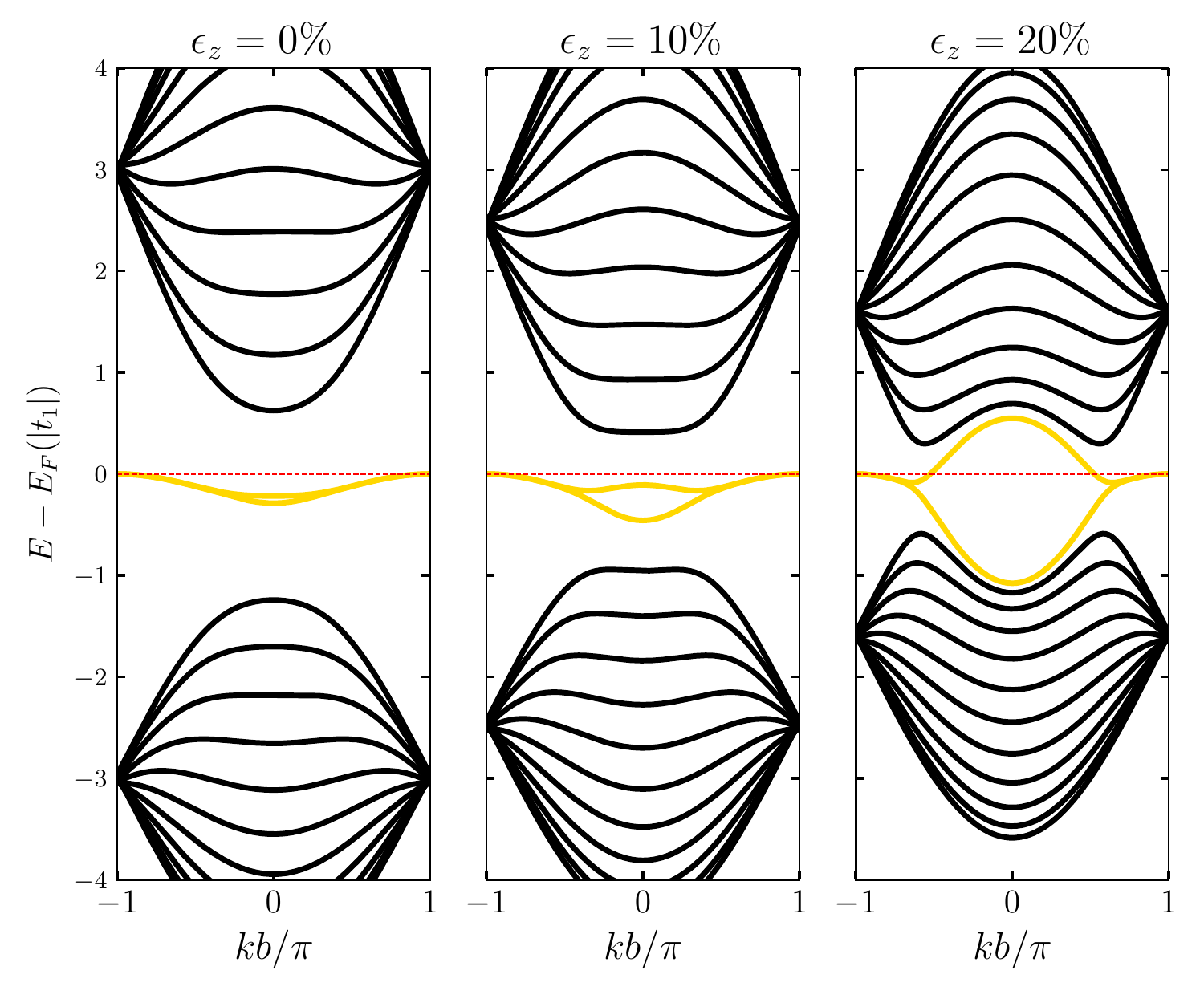} 
\caption{
Energy dispersion of ZPNR shown for three different strengths of tensile strain $\varepsilon_z=0.0\%,~10\%,~20\%$ in the absence of the Hubbard interaction $U$. The two quasi-flat edges, isolated from the bulk, are colored in gold. The width of the ribbon is the same as in Fig. \ref{fig4}. The horizontal red line marks the Fermi level.}
\label{fig8}
\end{figure}

Before exploring strain effects on the magnetic features, we briefly comment on the energy dispersion under strain. 
Figure~\ref{fig8} presents the energy dispersion for three different strengths of tensile strain
$\epsilon_z=0.0\%,~10\%,~20\%$ in the absence of the Hubbard interaction. It can be seen that the tensile strain has
a significant impact on the band structure: The edge modes are split, which is accompanied by a compression of the bulk bands. Even for strain $\epsilon_z=20\%$, the degeneracy of the edge modes at the BZ boundaries survives, while
one of the split levels crosses the Fermi energy at $k=0$. We also note that for strain $\epsilon_z=10\%$, 
the bulk band gets flattened, which is analogous to the strain-induced Landau Levels effects in graphene~\cite{Chang12,Vozmediano13,Roy13,Yang17}.
\begin{figure}
\includegraphics[width=1\columnwidth]{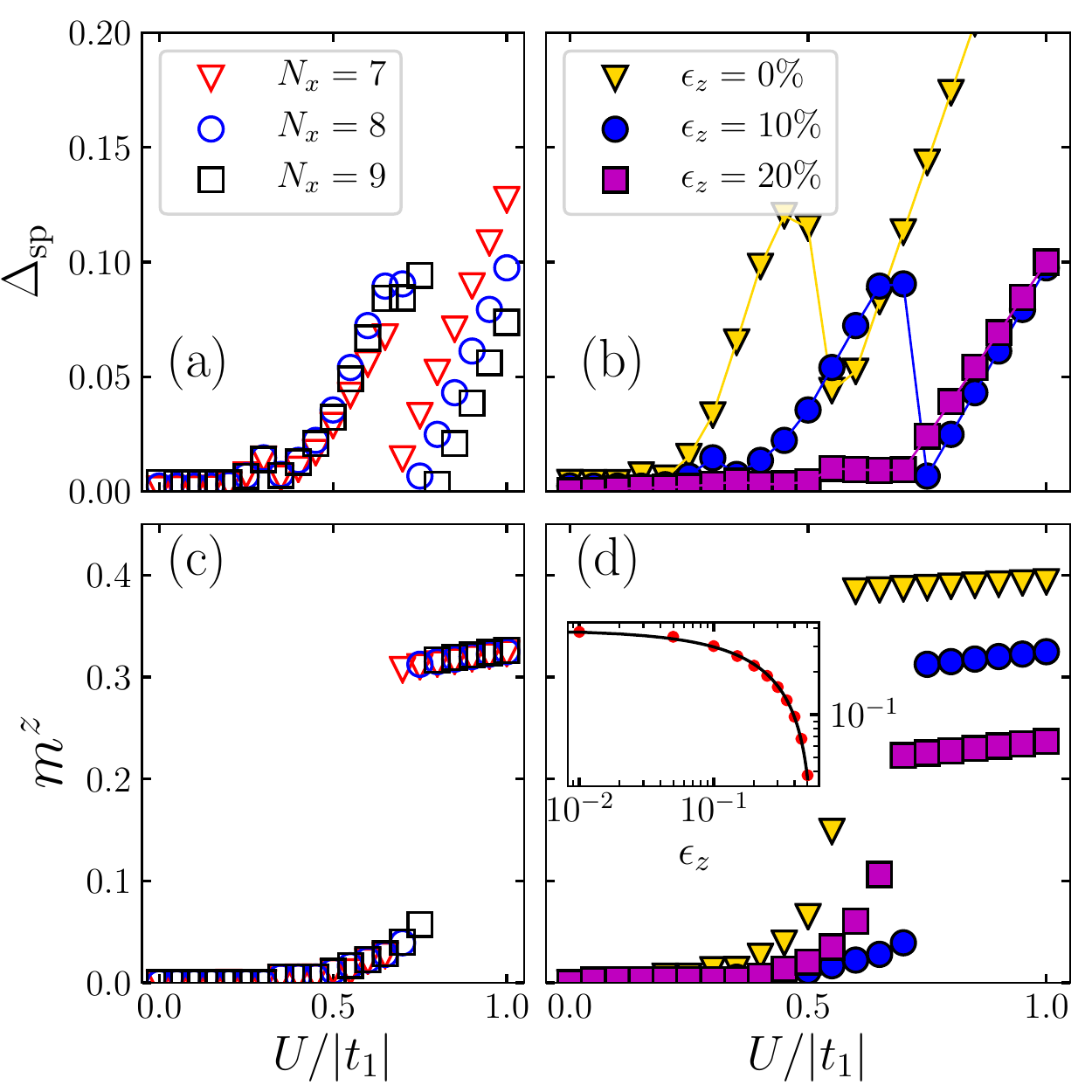} 
\caption{
 Panels (a) and (c)  show the gap and edge magnetization for different ribbon widths at fixed  tensile strain $\varepsilon_z=10\%$. In panels (b) and (d), we  fix the ribbon width, $N_x=7$, and show  three different strengths of tensile strain $\varepsilon_z=0.0\%,~10\%,~20\%$. The inset in panel (d) shows a  \textit{log}-\textit{log} plot of the $m^{z}$ evolution as a function of the tensile strain strength at $U/|t_1|=0.8$ (the line is a power-law fit).} 
\label{fig9}
\end{figure}
\par
We now explore the evolution of the gap and the magnetization as a function of the Hubbard interaction in the presence of tensile
strain. Figures~\ref{fig9} (a) and (c) show the results for different ribbon widths with fixed strain $\epsilon_z=10\%$.
Figures~\ref{fig9} (b) and (d) show the results  for  three different strengths of tensile strain $\epsilon_z=0.0\%,~10\%,~20\%$ with a fixed ribbon width $N_x=7$. One profound effect of tensile strain is the destruction of the intermediate IC phase. For $\epsilon_z=~20\%$, the IC phase has almost vanished. We furthermore notice that the second critical point, $U_{c2}$, shifts to a larger value. Thus, the tensile strain has a tremendous impact on edge magnetism. This can be understood by the following explanation: under tensile strain, the  $t_2$  and $t_4$ hopping parameters change more strongly than the others. As shown in Fig.~\ref{fig8}, the tensile strain splits the edge modes and increases their width. The increased bandwidth of these modes makes a larger interaction strength necessary to stabilize the AFM phase. Furthermore, as mentioned earlier, the $t_4$ hopping term is important for the shape of the edge mode and thus plays an essential role in stabilizing the IC phase.
Extrapolating the magnetization $m_z$ to larger values of tensile strain for $U/|t_1|=0.8$, shown in the inset of Fig.~\ref{fig9}(d),  
we find that magnetism should vanish at about $\epsilon_z=50\%$, which is much higher than the prediction by first principles in Ref~\cite{Du15}. We note that a strain of about $\epsilon_z=50\%$ is already big enough to destroy the whole structure of the edge modes. Thus, at the MFT level, we can conclude that a magnetic to nonmagnetic transition is not feasible.

\begin{figure}[b]
\includegraphics[width=1\columnwidth]{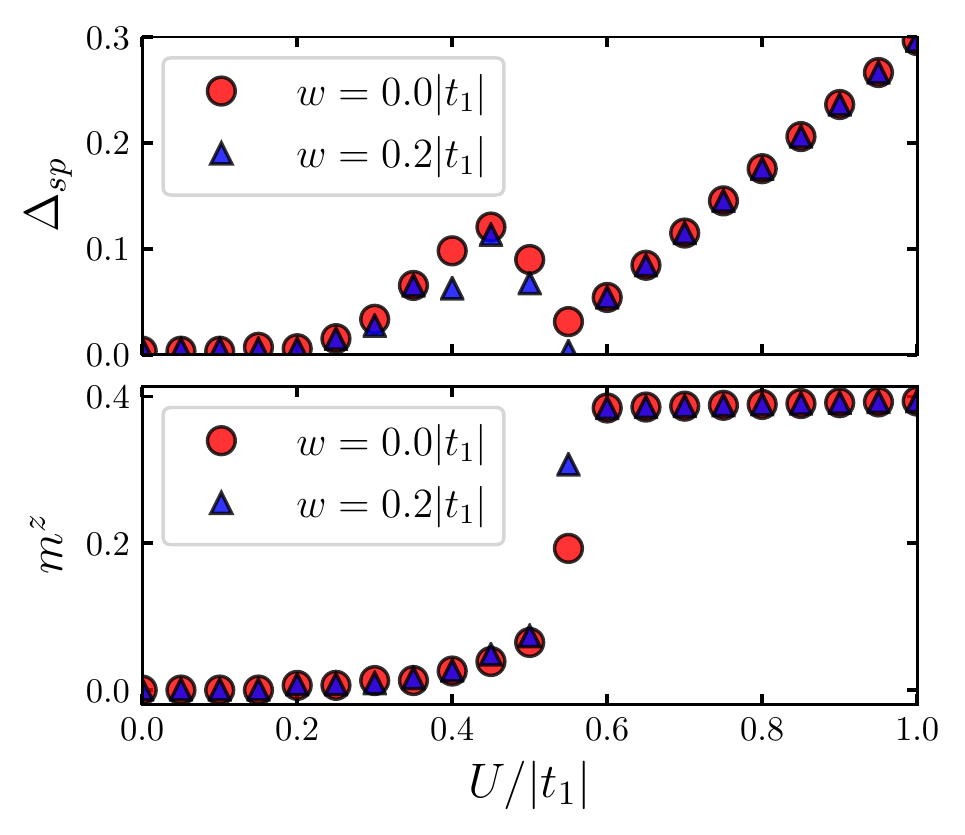} 
\caption{
Evolution of the single-particle gap (upper panel) and edge magnetization (lower panel) as a function of the Hubbard interaction $U$. Data for  clean and disorder cases $w_i/|t_1|=0.0,~0.3$ are shown. The width of the ribbon is the same as in Fig. \ref{fig4}.} 
\label{fig11}
\end{figure}
\subsection{Disorder effects}
To study the effects of disorder, we  include an additional term along the edges in the Hamiltonian $H_w=\sum_{i,\sigma}w_in_{i,\sigma}$, corresponding to non-magnetic disorder. $w_i$ is the strength of the disorder at site $i$, which is randomly chosen in the interval $[-w/2;w/2]$. Because the  translational symmetry along the y-direction is broken, we solve for the ground state in a finite cluster with $N_y=120$ using the periodic boundary condition. To be independent of a special configuration of the disorder, we average over $100$ different realizations. The influence of the edge disorder on the magnetic phases is presented in Fig.~\ref{fig11}.
 It can be seen that both, the gap and edge magnetization, are robust against disorder. We have not found any deviation in the saturation value of the edge magnetization in the gapped-AF(FM) phase. Moreover, the evolution of the gap also indicates the existence of the gapped-IC phase in the disordered system. This is in contrast to the result of strain in the preceding section. However, as mentioned before, the hopping parameters $t_2$ and $t_4$, play an essential role in stabilizing the edge states and its corresponding magnetic features. Thus, introducing Anderson type disorder will not destabilize these states.

\section{SUMMARY}
\label{sec4}
We have investigated the edge-state magnetic properties of black phosphorene nanoribbons using a tight-binding model with an electron-electron Hubbard interaction $U$.
 Our study aimed to explore the magnetic features of large clusters of black phosphorene nanoribbons for which numerically expensive techniques such as density functional theory and quantum Monte Carlo techniques are not feasible. Thus, to study the model, we have used a combination of static and dynamical mean-field theory (DMFT). 
While our calculations for large $U$ are in agreement with previous results, we find an {\em incommensurate} magnetic phase for weak interactions. Performing a detailed Fourier analysis of the magnetization evolution in the {\em incommensurate} (IC) phase, we find a second critical interaction $U_{c2}$  at which the IC phases changes to an antiferromagnetic (AFM) or ferromagnetic (FM) phase. 
Finally, we have analyzed the influence of strain and disorder on the magnetic properties. Our results show that while the IC phase is robust to Anderson type disorder, it is fragile against strain.

\begin{acknowledgments}
This work was supported by the Paris//Seine excellence initiative. J. Vahedi is also partially supported by Iran Science Elites Federation, Grant No.11/66332.  R.P. is supported by JSPS, KAKENHI Grant No. JP18K03511. 

\end{acknowledgments}
\appendix
\section{Unfolding of the Green's function}\label{Unfold}
When calculating the band structure for the long-range spin-density waves, we use extended unit cells including many layers in the y-direction. This yields a folded band structure, which is difficult to compare with the ferromagnetic or nonmagnetic state. We therefore unfold the band structure using the Green's function as described here.
When calculating the cluster Green's function with open boundary conditions in the x-direction, but including $k_y$ momentum dependence in the y-direction we obtain
\begin{equation}
    G^{k_y^\prime}_{y_1^\prime,y_2^\prime}(\omega)=\left(\omega+i\eta-H_{k_y^\prime}\right)^{-1}_{y_1^\prime,y_2^\prime},
\end{equation}
where $0\le y_{1,2}^\prime<N_y$ correspond to the $y$-component of different lattice sites in the unit cell  and $H_{k_y}$ is the Hamiltonian for the momentum $k_y$. Because we do not change the $x$-coordinate, we neglect it for convenience.
To calculate the unfolded Green's function, we need to calculate
\begin{equation}
    G_{k_y}(\omega)=\frac{1}{N}\sum_{y_1,y_2}\exp\left(-ik_y(y_1-y_2)\right)G_{y_1,y_2}(\omega)
\end{equation}
where $0\le y_{1,2}<N$ and $N$ the is number of lattice sites of the full lattice which includes $M$ unit cells with $N_y$ atoms, thus $N=MN_Y$.
We can calculate $G_{y_1,y_2}(\omega)$ from the cluster Green's function as
\begin{eqnarray}
    G_{y_1,y_2}(\omega)&=&\frac{1}{M}\sum_{k^\prime_y}  G^{k_y^\prime}_{y_1^\prime,y_2^\prime}(\omega)\exp\left( ik_y^\prime (n_1-n_2) \right)\\
       y_1&=&n_1N_y+y_1^\prime\\
    y_2&=&n_2N_y+y_2^\prime
\end{eqnarray}
We can now calculate the unfolded Green's function as
\begin{widetext}
\begin{eqnarray}
 G_{k_y}(\omega)&=&\frac{1}{N}\sum_{y_1,y_2}\exp\left(-ik_y(y_1-y_2)\right)\frac{1}{M}\sum_{k^\prime_y}  G^{k_y^\prime}_{y_1^\prime,y_2^\prime}(\omega)\exp\left( ik_y^\prime (n_1-n_2) \right)\nonumber\\
 &=&\frac{1}{NM}\sum_{n_1,n_2}\sum_{y_1^\prime,y_2^\prime}\sum_{k_y^\prime} \exp\left(-ik_y((n_1-n_2)N_y+y_1^\prime-y_2^\prime)\right)G^{k_y^\prime}_{y_1^\prime,y_2^\prime}(\omega)\exp\left( ik_y^\prime (n_1-n_2) \right)\nonumber\\
 &=&\frac{M^2}{NM}\sum_{y_1^\prime,y_2^\prime}\sum_{k^\prime_y} \delta_{N_yk_y,k_y^\prime} \exp\left(-ik_y(y_1^\prime-y_2^\prime)\right)G^{k_y^\prime}_{y_1^\prime,y_2^\prime}(\omega)\nonumber\nonumber\\
 &=&\frac{1}{N_y}\sum_{y_1^\prime,y_2^\prime}\exp\left(-ik_y(y_1^\prime-y_2^\prime)\right)G^{N_yk_y}_{y_1^\prime,y_2^\prime}(\omega)
\end{eqnarray}
\end{widetext}
Finally, we can calculate the spectral functions as shown in the main text as
\begin{equation}
    A_{k_y}(\omega)=-\frac{1}{\pi}\text{Im}\left(G_{k_y}(\omega)\right)
\end{equation}

\bibliography{Ref}

\end{document}